\documentclass{PoS}

\usepackage{amsmath}
\usepackage{multirow}

\newcommand{\bra}[1]{\langle #1|}
\newcommand{\ket}[1]{|#1\rangle}
\newcommand{\braket}[1]{\langle #1\rangle}
\newcommand{\Tr}[1]{\mathrm{Tr}\left[#1\right]}
\def\lsim{\mathrel{\rlap{\lower4pt\hbox{\hskip1pt$\sim$}}
    \raise1pt\hbox{$<$}}}                % less than or approx. symbol
\def\gsim{\mathrel{\rlap{\lower4pt\hbox{\hskip1pt$\sim$}}
    \raise1pt\hbox{$>$}}}                % greater than or approx. symbol

\title{The Spectral Structure of Correlator Matrices}

\ShortTitle{The Spectral Structure of Correlator Matrices}

\author{\speaker{Adam C. Lichtl}%
         \thanks{For the LHP Collaboration.}\\
        RBRC, Brookhaven National Laboratory\\
        E-mail: \email{alichtl@bnl.gov}}

\abstract{In lattice QCD spectrum calculations, it is desirable to obtain multiple excited state energies in each symmetry channel.  Typically, one constructs several interpolating operators for the symmetry channel of interest, forms the `correlator matrix' of all possible two-point functions, and uses the variational method to obtain as many energy levels as possible.  We present a detailed look at this last step, starting from a discussion of the symmetry properties and spectral structure of the correlator matrix.  We continue by motivating and describing the variational method, before discussing some conceptual and practical challenges concerning the light baryon sector.  We conclude by mentioning some alternate spectrum extraction methods currently under study.  Throughout, we attempt to quantify all approximations and assumptions, and we illustrate our points using a nucleon correlator matrix estimated on dynamical two-flavor lattice data.}

\FullConference{The XXV International Symposium on Lattice Field Theory\\
                 July 30 - August 4 2007\\
                 Regensburg, Germany}

\begin{document}

\section{Formalism}
 We will work with a $D\times D$ matrix of two-point functions, the so-called {\em correlator} matrix:
\begin{equation}
C^{(P)}_{ab}(\tau)\equiv \frac{1}{Z} \Tr{e^{-(\beta-\tau)H}\mathcal{O}^{(P)}_a e^{-\tau H}\overline{\mathcal{O}}^{(P)}_b},
\end{equation}
where $Z=e^{-\beta H}$ is the partition function, $\beta$ is the temporal extent of the lattice, $H$ is the Hamiltonian of the system, and the $D$ operators $\{\overline{\mathcal{O}}^{(P)}_a\}$ are designed to have:
\begin{itemize}
\item Zero three-momentum: $\vec{p}=0$,
\item Gauge invariance: singlet color structure,
\item Definite lattice spin~\cite{bib:colin_gt}: $G_1$, $H$, or $G_2$ (double-valued irreducible representations\footnote{The operators we use do not mix among the different lattice spin irreps.} of $O_h$),
\item Definite parity $P=\pm$, also denoted as: g = {\em gerade} (even) and u = {\em ungerade} (odd).
\end{itemize}
The operators carry baryon number $B=1$ and have been constructed to have the following behavior under charge conjugation, represented by the unitary operator $U_C$:
\begin{equation}
U_C \overline{\mathcal{O}}^{(P)}_a U_C^\dagger = U_C \overline{\mathcal{O}}^{(P)}_a U_C = \mathcal{O}^{(-P)}_a,\label{eqn:charge_conj}
\end{equation}
where we have dropped the dagger because $(U_C)^2=I$.  This behavior leads to the following relation between the positive- and negative-parity correlator matrices:
\begin{eqnarray}
C^{(P)}_{ab}(\tau) &=& \frac{1}{Z}\Tr{e^{-(\beta-\tau)H}\mathcal{O}^{(P)}_a e^{-\tau H}\overline{\mathcal{O}}^{(P)}_b},\\
&=&\frac{1}{Z}\Tr{e^{-(\beta-\tau)H}U_C U_C \mathcal{O}^{(P)}_a U_C U_C e^{-\tau H}\overline{\mathcal{O}}^{(P)}_b},\\
&=& \frac{1}{Z} \Tr{e^{-(\beta-\tau)H}U_C \mathcal{O}^{(P)}_a U_C  e^{-\tau H}U_C \overline{\mathcal{O}}^{(P)}_b U_C},\\
&=& \frac{1}{Z} \Tr{e^{-(\beta-\tau)H} \overline{\mathcal{O}}^{(-P)}_a e^{-\tau H}\mathcal{O}^{(-P)}_b },\\
&=&  \frac{1}{Z}\Tr{ e^{-\tau H}\mathcal{O}^{(-P)}_b e^{-(\beta-\tau)H} \overline{\mathcal{O}}^{(-P)}_a },\\
&=&  C^{(-P)}_{ba}(\beta-\tau),\label{eqn:parity_relation}
\end{eqnarray}
where we have used the cyclic property of the trace and the fact that charge conjugation is a symmetry of the theory:
$[H, U_C]=0$.  This relation is used to improve our statistics by averaging our numerical estimates of $C^{(P)}_{ab}(\tau)$ with those of $C^{(-P)}_{ba}(\beta-\tau)$. 

The system is defined in a finite volume and thus has a discrete spectrum.  Writing the trace as a sum over energy eigenstates $\{\ket{n}\}$ and inserting a second complete set of eigenstates $\{\ket{k}\}$ gives us the general spectral representation of the correlator matrix:
\begin{eqnarray}
C^{(P)}_{ab}(\tau) &=&\frac{1}{Z} \Tr{e^{-(\beta-\tau)H}\mathcal{O}^{(P)}_a e^{-\tau H}\overline{\mathcal{O}}^{(P)}_b},\\
&=& \frac{1}{Z}\sum_{n=0}^\infty \sum_{k=0}^\infty \bra{n}e^{-(\beta-\tau)H}\mathcal{O}^{(P)}_a e^{-\tau H}
\ket{k}\bra{k}\overline{\mathcal{O}}^{(P)}_b \ket{n},\\
&=& \frac{1}{Z}\sum_{n=0}^\infty \sum_{k=0}^\infty \bra{n}\mathcal{O}^{(P)}_a
\ket{k}\bra{k}\overline{\mathcal{O}}^{(P)}_b \ket{n}, e^{-E_k \tau}e^{-E_n(\beta-\tau)}.\label{eqn:full_spec}
\end{eqnarray}
In Eq.~\ref{eqn:full_spec}, we may freely divide a factor of $e^{-E_0\beta}$ out of the bottom and top, effectively canceling any zero-point energy of the theory: $H\ket{0}=0$.
Each of our baryon operators  $\overline{\mathcal{O}}^{(P)}_a$
creates a three-quark system of a given parity $P$ and annihilates
a three-antiquark system of the {\em same} parity $P$.  Unlike
bosons, a fermion and its antifermion have opposite intrinsic parities;
an antibaryon's energy is therefore the same as that of its corresponding {\em opposite-parity} baryon state, its so-called \textit{opposite-parity partner}.  Since
chiral symmetry is broken in QCD, the 
masses of the opposite-parity partners may differ.

Assuming throughout that $\bra{k}\overline{\mathcal{O}}^{(P)}_b \ket{n} \lsim \bra{k'}\overline{\mathcal{O}}^{(P)}_b \ket{n'}$ for $n, k \le n', k'$, the leading terms in the spectral representation involve the following non-vanishing  matrix elements:
\begin{itemize}
\item $\bra{k}\overline{\mathcal{O}}^{(P)}_a \ket{0}$ where $\ket{k}$ is a baryon state ($B=1$) having parity P and energy $E_k^{(P)}$,
\item $\bra{0}\overline{\mathcal{O}}^{(P)}_a \ket{n}$ where $\ket{n}$ is an antibaryon state ($B=-1$) having parity P and energy $E_n^{(-P)}$,
\item $\bra{BB}\overline{\mathcal{O}}^{(P)}_a \ket{B}$ where $\ket{B}$ has baryon number $B=1$ and $\ket{BB}$ has baryon number $B=2$,
\item $\bra{\bar{B}}\overline{\mathcal{O}}^{(P)}_a \ket{\bar{B}\bar{B}}$ where $\ket{\bar{B}}$ has baryon number $B=-1$ and $\ket{\bar{B}\bar{B}}$ has baryon number $B=-2$.
\end{itemize}
In the last two matrix elements above, a positive-parity operator will connect single- and multi-(anti)baryon states with the same parity, while a negative-parity operator will connect single and multi-(anti)baryon states with opposite parity.

Treating the largest terms explicitly, and using charge conjugation to convert the antibaryon states $\ket{n}$ into baryon states $\ket{k'}$ (the trace is invariant under charge conjugation) gives:
\begin{eqnarray}
C^{(P)}_{ab}(\tau) &=& \frac{1}{1+e^{-E_1 \beta}+\cdots} \sum_{k=1}^\infty \bra{0}\mathcal{O}^{(P)}_a
\ket{k}\bra{k}\overline{\mathcal{O}}^{(P)}_b \ket{0} e^{-E_k^{(P)} \tau} \nonumber\\
&+& \frac{1}{1+e^{-E_1 \beta}+\cdots}\sum_{k'=1}^\infty \bra{0}\mathcal{O}^{(-P)}_b\ket{k'}\bra{k'}\overline{\mathcal{O}}^{(-P)}_a \ket{0} e^{-E_{k'}^{(-P)}(\beta-\tau)}\nonumber\\
&+& \bra{1}\mathcal{O}\ket{1,1}\bra{1,1}\overline{\mathcal{O}} \ket{1} e^{- E_1\tau}e^{-E_1(\beta-\tau)}\nonumber\\
&+& \cdots.
\end{eqnarray}
where $E_1$ is the energy of the lightest baryon state $\ket{1}$ accessible by the action of our operators on the vacuum, regardless of parity.  We have approximated the energy of the lightest interacting $B=2$ state $\ket{1,1}$ as $E_1$, to be conservative.  Thus, all higher-order terms are suppressed by a factor of $e^{-E_1\beta}$, which is $O(10^{-6})$ in our studies.  At this point, we note that before neglecting terms, it is important to keep an eye on $e^{-E_k\tau}$ vs. $e^{-E_1\beta}$.  Dropping the higher-order terms gives us a spectral representation of the correlator matrix in terms of single-baryon states only:
\begin{eqnarray}
C^{(P)}_{ab}(\tau) &\approx&  \sum_{k=1}^\infty \bra{0}\mathcal{O}^{(P)}_a
\ket{k}\bra{k}\overline{\mathcal{O}}^{(P)}_b \ket{0} e^{-E_k^{(P)} \tau} \nonumber\\
&+& \sum_{k'=1}^\infty \bra{0}\mathcal{O}^{(-P)}_b\ket{k'}\bra{k'}\overline{\mathcal{O}}^{(-P)}_a \ket{0} e^{-E_{k'}^{(-P)}(\beta-\tau)},\label{eqn:approx}\\
&=&   \bra{0}\mathcal{O}^{(P)}_a e^{-H\tau}\overline{\mathcal{O}}^{(P)}_b \ket{0} 
+ \bra{0}\mathcal{O}^{(-P)}_b e^{-H(\beta-\tau)}\overline{\mathcal{O}}^{(-P)}_a \ket{0},\\
&\equiv& \braket{\mathcal{O}^{(P)}_a| e^{-H\tau} | \mathcal{O}^{(P)}_b} + \braket{\mathcal{O}^{(-P)}_b | e^{-H(\beta-\tau)} | \mathcal{O}^{(-P)}_a}.\label{eqn:correlator_long}
\end{eqnarray}
Notice that this approximation maintains the previously established relation (Eq.~\ref{eqn:parity_relation}): $C^{(P)}_{ab}(\tau)=C^{(-P)}_{ba}(\beta-\tau)$.  In the following, we will treat the approximation (Eq.~\ref{eqn:approx}) as exact and will refer to the two terms in Eq.~\ref{eqn:correlator_long} as the `forward' and `backward' parts of the correlator matrix, respectively.  For convenience, we define:
\begin{equation}
\widetilde{C}^{(P)}_{ab}(\tau)\equiv \braket{\mathcal{O}^{(P)}_a| e^{-H\tau} | \mathcal{O}^{(P)}_b},
\end{equation}
which allows us to write:
\begin{eqnarray}
C^{(P)}_{ab}(\tau)&=&\widetilde{C}^{(P)}_{ab}(\tau)+\widetilde{C}^{(-P)}_{ba}(\beta-\tau),\label{eqn:fwd_bkd1}\\
&=&\widetilde{C}^{(P)}_{ab}(\tau)+\widetilde{C}^{(-P)*}_{ab}(\beta-\tau).\label{eqn:fwd_bkd2}
\end{eqnarray}
 It is our ultimate goal to extract the spectrum $E_k^{(P)}, E_{k'}^{(-P)}$ for the various QCD hadron sectors.
 
 \section{The variational method}
Ideally, we would like to work with a very large temporal extent $\beta >> \tau$ such that:
\begin{equation}
C^{(P)}_{ab}(\tau)\stackrel{\beta>>\tau}{\to}\widetilde{C}^{(P)}_{ab}(\tau)= \braket{\mathcal{O}^{(P)}_a| e^{-H\tau} | \mathcal{O}^{(P)}_b},
\end{equation}
 in which case the problem reduces to simply diagonalizing the operator $e^{-H\tau}$ in the subspace defined by the trial basis $\{\ket{\mathcal{O}^{(P)}_a}\}$.  This can be accomplished with the variational method~\cite{bib:lichtl_dissertation, bib:luscher_variational, bib:dudek_charmonium_spectrum}.

We begin by refining the states in our trial basis by (formally) evolving them by some amount $\tau_0/2$ in imaginary time.  This serves to `relax' them onto the lower-lying energy states of interest:
\begin{eqnarray}
\ket{\mathcal{O}^{(P)}_a(\tau_0)} &\equiv& e^{-H\tau_0/2}\ket{\mathcal{O}^{(P)}_a},\\
&=& e^{-H\tau_0/2}\overline{\mathcal{O}}^{(P)}_a\ket{0},\\
&=&\sum_{k=1}^\infty \ket{k}\left[e^{-E_k\tau_0/2}\braket{k|\overline{\mathcal{O}}^{(P)}_a|0} \right],\label{eqn:trial}
\end{eqnarray}
where we have used the fact that $\braket{0|\overline{\mathcal{O}}^{(P)}_a|0} = 0$.  For sufficiently small $\tau_0$, the trial basis is linearly independent, but is not necessarily orthogonal or normalized:
\begin{equation}
\braket{\mathcal{O}^{(P)}_a(\tau_0)|\mathcal{O}^{(P)}_b(\tau_0)} = \widetilde{C}^{(P)}_{ab}(\tau_0).
\end{equation}
We can use the variational method to diagonalize the operator $e^{-H(\tau-\tau_0)}$ in our trial subspace by forming the linear combinations:
\begin{equation}
\ket{\Theta_i}\equiv \sum_{a=1}^D\ket{\mathcal{O}^{(P)}_a(\tau_0)}v_{ai},
\end{equation}
where the complex numbers $v_{ai}$ are the variational coefficients obtained by finding the extrema of:
\begin{equation}
\frac{\bra{\Theta_i}e^{-H(\tau-\tau_0)}\ket{\Theta_i}}{\braket{\Theta_i|{\Theta_i}}}=\frac{v_i^\dagger \widetilde{C}^{(P)}(\tau) v_i}{v_i^\dagger \widetilde{C}^{(P)}(\tau_0)v_i}.\label{eqn:extrema}
\end{equation}
Setting the derivative of Eq.~\ref{eqn:extrema} with respect to the $v_{ai}$ to zero yields a generalized eigenvalue problem whose solutions are the columns $v_i$ of the variational coefficient matrix $V$:
\begin{equation}
\widetilde{C}^{(P)}(\tau) v_i = \lambda_i \widetilde{C}^{(P)}(\tau_0)v_i.\label{eqn:gen_ev}
\end{equation}
The eigenvalues $\{\lambda_i\}$ appearing in Eq.~\ref{eqn:gen_ev} are known as {\em principal correlators}, and if all has gone well, they should asymptote in $\tau$ as $e^{-E^{(P)}_i(\tau-\tau_0)}$ while the $v_k$ maintain the orthogonality of the associated states $\{\ket{\Theta_i}\}$.  From the properties of the generalized eigenvalue problem, we can see that the metric of the vectors $v_i$ used to diagonalize $\widetilde{C}^{(P)}(\tau)$ at later times is fixed at the reference time $\tau_0$ by:
\begin{equation}
\braket{\Theta_i|{\Theta_j}} = v_i^\dagger \widetilde{C}^{(P)}(\tau_0) v_j = \delta_{ij}.\label{eqn:ortho_ev}
\end{equation}
It should be noted that the $v_{ai}$, $\lambda_i$, and $\ket{\Theta_i}$ all depend on both $\tau$ and $\tau_0$.

From the spectral expansion of the $D$ trial basis states in Eq.~\ref{eqn:trial}, we see that if we can choose $\tau_0$ large enough such that
 $\left[e^{-E_k\tau_0/2}\braket{k|\overline{\mathcal{O}}^{(P)}_a|0}\right]$ is negligible for $k > D$, then Eq.~\ref{eqn:ortho_ev} implies that:
\begin{equation}
\braket{k|{\Theta_j}} = \sum_{a=1}^D\braket{k|e^{-H\tau_0/2}\overline{\mathcal{O}}^{(P)}_a|0}v_{aj}  \propto \delta_{kj},\quad j,k \le D, \label{eqn:ortho}
\end{equation}
and that
\begin{equation}
\bra{\Theta_k}e^{-H(\tau-\tau_0)}\ket{\Theta_k}=v_k^\dagger \widetilde{C}^{(P)}(\tau)v_k = \lambda_k \approx e^{-E^{(P)}_k(\tau-\tau_0)}\label{eqn:asymp}
\end{equation}
occurs almost immediately for $\tau>\tau_0$.  If $\tau_0$ is not sufficiently large, it can be shown that the asymptotic form (Eq.~\ref{eqn:asymp}) is still reached, but more gradually.  In this case, it is possible that the signal will be lost in the noise before a reliable estimate of $E^{(P)}_k$ can be made.

We would like to choose $\tau_0$ to be large enough such that Eq.~\ref{eqn:ortho} holds to within statistical uncertainty, but are limited by three main factors.   First, we must estimate the correlator matrix numerically; the signal-to-noise ratio decays exponentially, leading to a noisy estimate of the trial basis for large $\tau_0$.  Second, as we increase $\tau_0$ towards $\beta/2$, the spectral terms in Eq.~\ref{eqn:approx} become exponentially small, and thus the rank of $C^{(P)}(\tau_0)$ will drop below $D$.  When this occurs, the condition number of the matrix will blow up because the trial basis will not have full rank, and the generalized eigenvalue problem will be numerically unstable.  This could be remedied by selecting some subset of operators (reducing $D$), but would incur the cost of throwing away information-containing entries of the original correlator matrix.  Finally, as we increase $\tau_0$, the previously neglected backward term $\widetilde{C}^{(-P)*}(\beta-\tau_0)$ will begin to contribute, with the particularly serious consequence that:
\begin{equation}
\braket{\mathcal{O}^{(P)}_a(\tau_0)|\mathcal{O}^{(P)}_b(\tau_0)} \neq \widetilde{C}^{(P)}_{ab}(\tau_0).
\end{equation}
This means, for example, that as we increase $\tau_0$ to reduce the positive-parity contamination, we will introduce negative-parity contamination.  These issues are not always troublesome in practice, but we have found  that they are indeed present in our nucleon spectrum study.

\clearpage

\section{Eigenvalue behavior}
\begin{figure}[h]
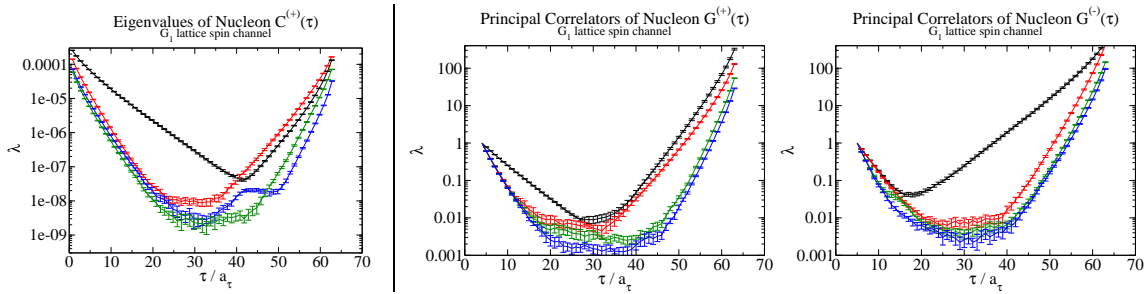

\begin{center}
\begin{minipage}{0.32\textwidth}
\includegraphics[width=\textwidth]{G1g_lambdas_b}
\end{minipage}
\hfill
\vline
\hfill
\begin{minipage}{0.32\textwidth}
\includegraphics[width=\textwidth]{princorr_G1g}
\end{minipage}
\begin{minipage}{0.32\textwidth}
\includegraphics[width=\textwidth]{princorr_G1u}
\end{minipage}
\end{center}
\caption{These three plots are associated with a $16\times16$ nucleon correlation matrix $C^{(P)}(\tau)$ in the $G_1$ lattice spin channel.
\textbf{Left:} the largest four eigenvalues of the positive parity matrix.  We see crossing and transient behavior due to the interference of the forward and backward parts of the correlator matrix.  Because our negative parity operators have a fixed relation to our positive parity operators, we do not have enough variational coefficients in this technique to separately isolate the positive- and negative-parity contributions when they are both present.  \textbf{Center and right:} the largest four principal correlators for the positive- and negative-parity correlator matrices, using $\tau_0 = 4a_\tau$ and $\tau_0=5a_\tau$, respectively.  In the center plot we see suspected crossing behavior around $\tau=30a_\tau$ and again around $\tau=45a_\tau$.  In the right plot we see significant opposite-parity partner contamination (from the proton) starting around $\tau=10a_\tau$.\label{fig:eigenvalues}}
\end{figure}

It is illuminating to examine the eigenspectrum (and rank) of an actual lattice QCD correlator matrix (Fig.~\ref{fig:eigenvalues}) as a function of $\tau$ before applying the variational method.  Here we consider a correlator matrix defined using $D=16$ previously designed~\cite{bib:colin_gt, bib:lichtl_dissertation, bib:lichtl_smear}
 nucleon operators (isospin $I=1/2$, $I_3=+1/2$), and estimated from 336 LHPC\footnote{Special thanks to Robert Edwards and Balint Jo\'o from Jefferson Lab.} two-flavor anisotropic Wilson lattice configurations having the following parameters:\\
\phantom{\qquad\qquad} $\bullet$ Lattice dimensions: $24^3\times 64$,\\
\phantom{\qquad\qquad} $\bullet$ Spatial lattice spacing: $a_s \approx 0.11$ fm,\\
\phantom{\qquad\qquad} $\bullet$ Anisotropy: $a_s/a_\tau \approx 3.0$,\\
\phantom{\qquad\qquad} $\bullet$ Pion mass: $m_\pi \approx 400$ MeV.

To find the $D=16$ eigenvalues $\lambda_j$ as functions of $\tau$, we solve $C^{(P)}(\tau)v_j(\tau) = \lambda_j(\tau)v_j(\tau)$ at each $\tau$.  The largest four $\lambda_j(\tau)$'s, corresponding asymptotically to the lowest four energy levels, are plotted on the left in Fig.~\ref{fig:eigenvalues}.  The most prominent feature of this plot is the apparent crossing and transient kinking of the eigenvalue flow with $\tau$.  The blue curve in particular exhibits interesting behavior:  it falls as $\exp[-E^{(+)}_2\tau]$ for early $\tau$, then becomes contaminated with the backward ground state $\exp[-E^{(-)}_1(\beta-\tau)]$ at intermediate $\tau$, and finally is forced (by the higher eigenvalues) to asymptote as $\exp[-E^{(-)}_4(\beta-\tau)]$. 
We may understand this qualitatively by the fact that the negative-parity operators excite the negative-parity energy levels differently than the positive-parity operators excite the positive-parity energy levels.  In other words: the coefficients diagonalizing $C^{(P)}(\tau)$ for $\tau<<\beta/2$ are largely independent of the coefficients diagonalizing $C^{(P)}(\tau)$ for $\tau >> \beta/2$.  Unfortunately, our variationally-determined negative-parity operators are completely determined by our variationally-determined positive-parity operators. 

To understand this behavior, we look at the spectral structure of $\lambda_j(\tau)$:
\begin{eqnarray}
\lambda_j(\tau) = v_j^\dagger(\tau) C^{(P)}(\tau)v_j(\tau) &=& v_j^\dagger(\tau)\widetilde{C}^{(P)}(\tau)v_j(\tau)+v_j^\dagger(\tau) \widetilde{C}^{(-P)*}(\beta-\tau)v_j(\tau),\\
&=& v_j^\dagger(\tau)\widetilde{C}^{(P)}(\tau)v_j(\tau)+v_j^T(\tau) \widetilde{C}^{(-P)}(\beta-\tau)v_j^*(\tau),\\
&\equiv& \braket{\Theta^{(P)}_j|e^{-H\tau}|\Theta^{(P)}_j} + \braket{\Theta^{(-P)}_j|e^{-H(\beta-\tau)}|\Theta^{(-P)}_j},
\end{eqnarray}
where we have introduced the states $\ket{\Theta^{(P)}_j}$ created by linear combinations of our basis operators:
\begin{equation}
\overline{\Theta}^{(P)}_j \equiv \sum_{a=1}^D \overline{\mathcal{O}}^{(P)}_a v_{aj}, \qquad
\overline{\Theta}^{(-P)}_j \equiv \sum_{a=1}^D \overline{\mathcal{O}}^{(-P)}_a v^*_{aj}.
\end{equation}
The variationally-determined $\overline{\Theta}^{(P)}_j$ operators behave as follows under charge conjugation (c.f. Eq.~\ref{eqn:charge_conj}):
\begin{equation}
U_C \overline{\Theta}^{(P)}_j U_C = \sum_{a=1}^D U_C \overline{\mathcal{O}}^{(P)}_a U_C v_{aj}= \sum_{a=1}^D \mathcal{O}^{(-P)}_a v_{aj} = \Theta^{(-P)}_j,
\end{equation}
and  thus we see that a brute-force application of the variational method to a correlator matrix having significant contributions from both forward and backward components will be plagued with contamination because we are using only one set of variational coefficients $v_{aj}$ to diagonalize two matrices  $\widetilde{C}^{(P)}(\tau)$ and  $\widetilde{C}^{(-P)*}(\beta-\tau)$, which separately have different eigenvectors.

\section{Outlook}

We are currently examining two alternate techniques for the extraction of the spectrum from a correlator matrix.  First, we are considering the possibility of  introducing more coefficients by taking linear combinations of the elements of $C^{(P)}(\tau)$ and $C^{(-P)}(\beta-\tau)$.  This may allow us to better decompose $C^{(P)}(\tau)$ by breaking the charge conjugation relationship of our variationally determined operators.  A second approach would be to attempt full matrix fits of increasing rank $M$ to the reduced matrix $R_{ij}(\tau)=v^\dagger_i C(\tau) v_j$, where the $v_i$'s are variationally determined at some fixed time $\tau^*$ and only $M$ of the column vectors $v_i$ are used ($i, j = 1,\ldots,M$) .
We expect continuing progress towards producing reliable determinations of the various excited baryon spectra predicted by QCD.
 
This research is supported by NSF grant PHY 0653315, and numerical calculations were performed using computing resources provided by Jefferson Lab and the RIKEN BNL Research Center (RBRC) at Brookhaven Lab.  Additional resources and travel support to the XXV International Symposium on Lattice Field Theory were provided by the RBRC.  The author would also like to thank J. Dudek, R. Edwards, C. Morningstar, M. Peardon,  and D. Richards for many fruitful discussions relating to this topic.

\end{document}